\documentclass[aps,prd,twocolumn,longbibliography,nofootinbib,floatfix]{revtex4-2}

\usepackage{graphicx}
\usepackage{xcolor}
\usepackage{mathrsfs,bbm}
\usepackage{amsmath,amsfonts,amssymb}
\usepackage{graphicx}
\usepackage{slashed}
\usepackage{multirow}
\usepackage{tensor}
\usepackage{hyperref}
\usepackage{url}
\usepackage[normalem]{ulem}
\usepackage[utf8]{inputenc}
\usepackage{orcidlink}

\allowdisplaybreaks

\def\p{\partial}

\begin{document}

\title{3D Evolution of a Good-Bad-Ugly-F Model on Compactified
  Hyperboloidal Slices}

\author{Christian Peterson$^{1}$\orcidlink{0000-0003-4842-1368}}
\author{Shalabh Gautam$^{2}$\orcidlink{0000-0003-2230-3988}}
\author{In\^es Rainho$^{1,3}$\orcidlink{0000-0003-4937-0638}}
\author{Alex Va\~n\'o-Vi\~nuales$^{1}$\orcidlink{0000-0002-8589-006X}}
\author{David Hilditch$^{1}$\orcidlink{0000-0001-9960-5293}}

\affiliation{${}^1$CENTRA, Departamento de F\'isica, Instituto
  Superior T\'ecnico IST, Universidade de Lisboa UL, Avenida Rovisco
  Pais 1, 1049-001 Lisboa, Portugal \\
  ${}^2$International Centre for Theoretical Sciences (ICTS), Survey
  No. 151, Shivakote, Hesaraghatta Hobli, Bengaluru - 560 089, India \\
  ${}^3$Departamento de Astronomía y Astrofísica, Universitat de
  València, Dr. Moliner 50, 46100, Burjassot (València), Spain}

\date{\today}

\begin{abstract}
  The \textit{Good-Bad-Ugly-F} model is a system of semi-linear wave
  equations that mimics the asymptotic form of the Einstein field
  equations in generalized harmonic gauge with specific constraint
  damping and suitable gauge source functions. These constraint
  additions and gauge source functions eliminate logarithmic
  divergences appearing at the leading order in the asymptotic
  expansion of the metric components. In this work, as a step towards
  using compactified hyperboloidal slices in numerical relativity, we
  evolve this model numerically in spherical symmetry, axisymmetry and
  full 3d on such hyperboloidal slices. Promising numerical results
  are found in all cases. Our results show that nonlinear systems of
  wave equations with the asymptotics of the Einstein field equations
  in the above form can be reliably captured within hyperboloidal
  numerical evolution without assuming symmetry.
\end{abstract}

\maketitle

\section{Introduction}
\label{Section:introduction}

The computation of gravitational waves at future null
infinity,~$\mathscr{I}^+$, is arguably the most important deliverable
from a numerical relativity~(NR) simulation of a coalescing binary of
compact objects in an asymptotically flat spacetime. Despite huge
progress, there is no first principles solution to this problem so
far. The main issue is that Cauchy evolution is the most widely used
approach in~NR, and since Cauchy slices have to be truncated to permit
practical evolutions, post-processing methods for extracting the
waveform out of the numerical domain have to be used to get the
desired signal. The two classic examples of this are direct
extrapolation and Cauchy-characteristic extraction, which are the most
widely used gravitational wave extraction methods in NR
codes~\cite{BisGomLeh97a, BisGomLeh97, ZloGomHus03, HanSzi14,
  BarMoxSch19, MoxSchTeu21}. There is an ongoing effort to
include~$\mathscr{I}^+$ directly in the computational domain via
Cauchy-characteristic matching. In this approach the wavezone is
foliated using compactified null slices and matched to standard Cauchy
hypersurfaces in the interior. See~\cite{Win12} for a review
and~\cite{GiaHilZil20, GiaBisHil21} for recent work on the
well-posedness of general relativity (GR) in the single-null gauges
typically employed by formulations for both characteristic extraction
and matching.

An alternative strategy to reach null-infinity numerically is to
foliate spacetime using compactified hyperboloidal slices~\cite{Fri81,
  Fri81a, Fra98, Fra98a, Fra04, MonRin08, Zen08, BarSarBuc11,
  VanHusHil14, VanHus14, Van15, VanHus17}. Like Cauchy hypersurfaces
these are everywhere spacelike, but terminate at~$\mathscr{I}^+$
instead of spatial infinity,~$i^0$. As~$\mathscr{I}^+$ is an infinite
distance apart from the evolution region, we need to compactify an
outgoing radial coordinate. A nice property of hyperboloidal slices is
that an outgoing solution to the wave equation, which serves as a
fundamental model for all systems with wavelike solutions, oscillates
just a finite number of times before reaching~$\mathscr{I}^+$. In
contrast, on Cauchy slices an infinite number of oscillations
transpire as a wave propagates out. In this sense, hyperboloidal
slices are adapted to resolve outgoing waves just as outgoing
characteristic slices. Consequently, one can expect to resolve
outgoing waves on hyperboloidal slices numerically with finite
resolution~\cite{Zen10}. The price paid is that incoming waves are
poorly resolved, but this is an acceptable loss because there should
be very little incoming radiation content from near~$\mathscr{I}^+$ in
the scenarios we are ultimately interested in computing.

In this paper, we use a foliation of Minkowski spacetime by
hyperboloidal slices to evolve a system of wave equations called the
\textit{Good-Bad-Ugly-F}~(GBUF) system. The work is a direct
continuation of our previous studies~\cite{GasGauHil19, GauVanHil21}.
In this earlier work, we studied the \textit{Good-Bad-Ugly} (GBU)
system. The GBU system mimics the Einstein Field Equations~(EFEs) in
the asymptotic future null directions in harmonic gauge by ignoring
their tensorial nature and discarding (many) sufficiently rapidly
decaying pieces of the solution in the equations of motion. The GBU
model consists of the equations
\begin{align}\label{GBU_model} \Box \, g = 0 \, , \quad \Box \, b =
  (\p_T \, g)^2 \, , \quad \Box \, u = \frac{2}{\sqrt{1+R^2}} \, \p_T
  u \, ,
\end{align}
where~$g$,~$b$ and~$u$ represent the good, bad and ugly fields
respectively. The numerical treatment of the homogeneous wave equation
on compactified hyperboloidal slices is now fairly standard. The key
interest in studying such models is instead in understanding the
effect that the inhomogeneities, whether linear or nonlinear, have on
the asymptotics of the fields, and how (or even if) these may be
treated numerically if they permit either fast or only very weak decay
in~$R$. For instance, the linear term on the right hand side of
the~$u$ equation suppresses the associated radiation field, which
leads to faster decay than that of solutions to the wave equation,
whereas the~``$gb$'' sector of the model satisfies not the classical
null condition~\cite{Kla80}, but rather the weak null
condition~\cite{LinRod03}. Consequently, near null-infinity, the bad
field encounters an obstruction to decay and, even starting from
initial data of compact support, solutions admit only a slow
asymptotic expansion of the form~\cite{LinRod03}
\begin{align}
  b \sim \frac{\log R}{R}B_1(T-R)+\frac{1}{R}B_2(T-R) \,. 
\end{align}
The logarithmic term here is problematic for numerics when one wants
to resolve the field asymptotically~\cite{GasGauHil19}. By analogy,
this suggests that plain harmonic gauge is not ideally suited for
hyperboloidal evolution. Inspired by the Generalized Harmonic
Gauge~(GHG) formulation of general relativity (GR) we therefore modify
the system~\eqref{GBU_model} by adding a variable~$f$ to the system
whose equation of motion we are free to choose. Inspired
by~\cite{LinSzi09}, we follow the general strategy
of~\cite{DuaFenGasHil21,DuaFenGas22,DuaFenGasHil22a}, to make an
appropriate choice of~$f$ which cures the behavior of the bad field
near~$\mathscr{I}^+$, resulting in asymptotics that are simpler to
deal with numerically. With these choices fixed, we implement the
model in a stand-alone spherical code and in the~3d
NRPy+~\cite{RucEtiBau18} infrastructure. Numerical results from both
implementations are found to be compatible.

The paper is organized as follows. In
section~\ref{GBUF_model_section}, we introduce the~GBUF model in
detail and describe its asymptotic properties. We also perform a first
order reduction of the system using radial characteristic variables,
introduce appropriate rescalings to regularize the equations
at~$\mathscr{I}^+$, and then present the limiting equations satisfied
at~$\mathscr{I}^+$, so as to show the equations in their final form to
be implemented. Section~\ref{numerics} continues with a presentation
of the details of the spherical and~NRPy+ implementations and their
respective numerical results. We demonstrate our spherically symmetric
results via two different choices of variables and numerical
schemes. The first one is the standard Evans method~\cite{Eva84} as
applied in~\cite{GasGauHil19}. The second one is similar to the
summation-by-parts~(SBP) method, as derived in~\cite{GauVanHil21}. We
close in section~\ref{conclusions} with a discussion of, and
conclusions from, the present work.

\section{The GBUF model}
\label{GBUF_model_section}

In this paper we generalize the GBU system of
equation~\eqref{GBU_model} to include a field~$f$. This system, the
\textit{Good-Bad-Ugly-F} model, serves as a model for GR in GHG,
rather than pure harmonic gauge. Specifically, an additional term is
added to the bad equation that mimics a part of the gauge source
terms, present in the~EFEs. The resulting system takes the form
\begin{align}\label{GBUF_model}
  \Box \, g = 0 \, , \quad \Box \, b
  = \frac{1}{\chi}\p_T f + (\p_T \, g)^2 \, ,
  \quad \Box \, u = \frac{2}{\chi} \, \p_T u \, ,
\end{align}
where~$g$,~$b$ and~$u$ represent the good, the bad and the ugly
fields, respectively, as described in~\cite{GasGauHil19},
and~$\chi = \sqrt{1+R^2}$ so that~$\chi \sim 1$ near the origin
and~$\chi \sim R$ for large~$R$. Here,~$f$ plays the role of the gauge
source function in~GHG. In~\cite{DuaFenGas22} it has been seen that
terms of this type can be used to regularize some of the equations
at~$\mathscr{I}^+$. The game is to make a specific choice of equation
of motion for~$f$ that achieves this goal. Before making such a
choice, we observe that~$f$ should fall-off at least like~$1/R$
towards~$\mathscr{I}^+$. Clearly, the GBU model corresponds to the
choice~$\p_T f \equiv 0$, which has been seen to lead to logarithmic
divergences in the bad field. In this paper, we take the equation of
motion for~$f$ to be
\begin{align}\label{GBUF_F}
\Box \, f = \frac{2}{\chi}\p_T f + 2(\p_T \, g)^2 \, .
\end{align}
Note that the first term on the right hand side of the above equation
gives it the form of the ugly wave equation, which does not radiate
towards~$\mathscr{I}^+$. On the other hand, the nonlinear term falls
off like~$1/R^2$, so it can be seen that this term determines the
fall-off of~$f$
towards~$\mathscr{I}^+$~\cite{LinRod03,DuaFenGasHil21}. With this
particular choice, the asymptotic expansion of the fields (within a
large class of initial data) is~\cite{DuaFenGas22}
\begin{align}\label{decay}
&g \sim \frac{G_1(T-R)}{R} \,, \nonumber \\ 
&b \sim \frac{B_1(T-R)}{R} \,,  \\
&u \sim \frac{U_1}{R} = \frac{m_u}{R} \, ,  \nonumber \\
&f \sim \frac{F_1(T-R)}{R} \,.  \nonumber
\end{align}
where~$G_1$ and~$F_1$ are related through~$F_1'(T-R)=-G_1'(T-R)^2/2$.
If we therefore rescale the fields by~$\chi$, they become~$O(1)$ all
the way towards~$\mathscr{I}^+$ and we can cleanly extract the
radiation field asymptotically. In contrast to the good or bad fields,
the leading order term, or the ``mass" term~$m_u$, in the ugly field
does not depend on advanced or retarded time. In other words the
radiation field associated with~$u$ is trivial. In this work, we
consider initial data~(ID) with~$m_u \equiv 0$, so that in fact
the~$u$ field decays one power~of~$R$ faster than the other
fields. Evolving~ID with a non-vanishing mass term with this model
could be achieved with a straightforward change of variables.

\subsection{First order reduction and rescaling}
\label{FT1S}

We now present the equations of motion in the form that will be
implemented numerically. We first make a naive first-order
reduction~(FOR) of the second-order equations~\eqref{GBUF_model}
and~\eqref{GBUF_F} in terms of their null and angular derivatives.
Writing~$\psi$ to represent any of the~$g$,~$b$,~$u$ or~$f$ fields, we
define FOR variables by
\begin{align}\label{FOR}
  \psi^+ = \p_T \psi + \p_R \psi , \quad \psi^- = \p_T \psi - \p_R \psi ,
  \quad \psi_A=\hat{\Theta}_A\psi \, ,
\end{align}
where~$A=\{\theta, \phi \}$
and~$\hat{\Theta}_A=\{ \p_{\theta},(1/\sin \theta) \p_\phi \}$.

The utility of this particular choice of~FOR variables is that, for a
large class of initial data, their fall-off rates
towards~$\mathscr{I}^+$ form a clean hierarchy. Knowledge of these
fall-off rates is then used for rescaling the respective variables
such that the resulting equations in hyperboloidal coordinates are
regular (as far as is possible) at~$\mathscr{I}^+$. One can see
from~\eqref{decay} that, with the exception of the ugly fields, the
outgoing characteristic variables~$\psi^-$ fall-off like~$1/R$ whereas
the ingoing ones,~$\psi^+$, decay like~$1/R^2$
towards~$\mathscr{I}^+$. The outgoing characteristic field associated
with~$u$ has instead faster decay like~$1/R^2$. The angular
derivatives,~$\psi_A$, fall-off like~$1/R$. The rescaled~FOR fields we
work with are
\begin{align}\label{Rescaled_FOR}
 \Psi &\equiv \chi \psi \,, \nonumber  \\
 \Psi^- &\equiv \chi \psi^- \,, \nonumber  \\
 \Psi^+ &\equiv \chi(\p_T+\p_R)(\chi\psi) \,, \\
 \Psi_A &\equiv \chi \psi_A \,. \nonumber 
\end{align}
Various alternative choices of reduction variables that capture the
stratification in decay rates are possible. Observe that despite the
fact that incoming null derivatives of the~$u$ field decay
like~$O(R^{-2})$, we rescale only by a single power of~$R$. This is to
permit the treatment of source terms that decay at best
like~$O(R^{-3})$, which appear for instance in GR proper.

Since we perform a complete reduction of the equations, the~FOR
variables have associated with them FOR constraints. In terms of the
rescaled variables these constraints read
\begin{align} \label{constraints_uppercase}
  &\p_R\Psi + \Psi^- -\frac{\chi'}{\chi}\Psi -\frac{1}{\chi}\Psi^+ = 0 \,, \nonumber \\
  &\hat{\Theta}_A \Psi - \Psi_A = 0  \, ,
\end{align}
where~$\chi' \equiv d\chi/dR$. If a solution of the FOR satisfies
these constraints, it can be unambiguously mapped to a solution of the
original second order equations. In numerical applications the
reduction constraints can be violated due to either a poor choice of
initial data, or by numerical error that should converge away with
resolution. As in any free evolution setup, we monitor the constraints
during evolution.

\subsection{Hyperboloidal coordinates and compactification}

The next step is to introduce a coordinate system that
maps~$\mathscr{I}^+$ on to a finite numerical grid in such a way that
outgoing radiation is well-resolved. For this purpose, we introduce
hyperboloidal slices as described in~\cite{Zen07a, Zen10, ZenKid10,
  GasGauHil19, GauVanHil21}. These slices use a hyperboloidal time
coordinate and a compactified radial coordinate defined on the level
sets of hyperboloidal time, denoted respectively by~$(t,r)$ and
defined as
\begin{align*}
t = T - H(R) \, , \; R = R(r) \, .
\end{align*}
Demanding that~$dH/dR<1$ for all~$R$ and~$dH/dR \rightarrow 1$ (fast
enough) as~$R\rightarrow \infty$, the level sets of~$t$ are spacelike
everywhere but reach~$\mathscr{I}^+$. The simplest example of such a
height function is given by~$H(R)=\sqrt{1+R^2}$, which explains
why~$t$ is called hyperboloidal time. To compactify the radial
coordinate we define~$R(r)=r/\Omega(r)$,
with~$\Omega(r)=1-r^2/r_{\mathscr{I}}^2$. Note that~$\mathscr{I}^+$ is
mapped to~$r = r_\mathscr{I}$, and for simplicity we
take~$r_{\mathscr{I}} = 1$.

We define the differential
operator~$\dot{\Box}_p \psi \equiv \Box \psi +\frac{2p}{\chi} \p_T
\psi$, so that each of the~GBUF wave equations has the
form~$\dot{\Box}_p \psi = S_{\psi}$, where $p=0$ for good and bad
equations and $p=1$ for the ugly and f ones. We work with the
particular choice~$H(R) = R - r(R)$. As discussed
in~\cite{GauVanHil21}, combining this choice of height function with
the compactification above results in a foliation of limited
regularity at the origin, which in practice does not appear to cause
leading-order problems for our second order accurate discretization at
the resolutions we have used. In these new coordinates, the equations
take the form
\begin{align*}
\p_t \Psi &= \frac{\Psi^+}{2\chi} +\frac{\Psi^-}{2} -\frac{\chi' \Psi}{2\chi} , \\
\p_t \Psi^+ &=\frac{1}{2R'-1} \left[ \left( \frac{R'}{R} -\frac{\chi'}{2\chi} -\frac{\chi'R'}{\chi} \right)\Psi^+  +\p_r\Psi^+ \right. \\
	& \left. +\left( -\frac{\chi R'}{R} -\frac{1}{2}\chi' +\chi' R' \right)\Psi^-  -\chi'\p_r\Psi \right. \\
	& \left. +\left( -\frac{\chi' R'}{R} +\frac{(\chi')^2}{2\chi}  +\frac{(\chi')^2R'}{\chi}  -\chi'' R' \right)\Psi \right. \\
& \left. +\frac{\cot(\theta)\chi R'}{R^2}\Psi_{\theta} +\frac{\chi R'}{R^2}\p_{\theta}\Psi_{\theta} +\frac{\chi R'}{\sin(\theta)R^2}\p_{\phi}\Psi_{\phi} \right. \\
& \left. -p\frac{R'\Psi^+}{\chi}  -p R'\Psi^-  +p\frac{\chi'R'}{\chi}\Psi  +\mathcal{S}_\Psi   \right] , \\
\p_t \Psi^- &= \frac{R'\Psi^+}{R\chi} -\left( \frac{R'}{R} -\frac{\chi'R'}{\chi}\right)\Psi^- -\p_r\Psi^- -\frac{\chi'R'\Psi}{R\chi} \\
& +\frac{\cot(\theta)R'}{R^2}\Psi_{\theta} +\frac{R'}{R^2}\p_{\theta}\Psi_{\theta} +\frac{R'}{\sin(\theta)R^2}\p_{\phi}\Psi_{\phi} \\
& -p\frac{R'\Psi^+}{\chi^2}  -p\frac{R'\Psi^-}{\chi} +p\frac{\chi'R'\Psi}{\chi^2}  +\frac{1}{\chi}\mathcal{S}_\Psi  , \\
\p_t \Psi_A &= -\frac{\chi'}{2\chi}\Psi_A +\frac{1}{2\chi}\hat{\Theta}_A\Psi^+ +\frac{1}{2}\hat{\Theta}_A\Psi^- ,
\end{align*}
where
\begin{align*}
  \mathcal{S}_B &= -\frac{R'}{4}\big( \frac{1}{\chi}G^+ +G^- -\frac{\chi'}{\chi}G \big)^2
                  -\frac{R'}{2\chi}F^+\\
                &\quad-\frac{1}{2}R'F^-  +\frac{\chi'R'}{2\chi}F \,, \\
  \mathcal{S}_F &=
                  -\frac{R'}{2}\big( \frac{1}{\chi}G^+ +G^- -\frac{\chi'}{\chi}G \big)^2 \,, \\
 \mathcal{S}_G &=  \mathcal{S}_U = 0 \,. 
\end{align*}
This is the system we evolve numerically, both in spherical symmetry
and full~3d. The constraints~\eqref{constraints_uppercase} in terms of
hyperboloidal coordinates read
\begin{align*}
  &\frac{1}{2R'-1}\left(2 \, \p_r\Psi + \Psi^-
    -\frac{\chi'}{\chi}\Psi \right) - \frac{1}{\chi}\Psi^+ = 0 \, ,\\
 &\hat{\Theta}_A \Psi - \Psi_A = 0  \, .
\end{align*}

Some terms in the previous system of equations have coefficients that
diverge asymptotically, but are multiplied by parts of the solution
that decay fast enough that the product takes a finite limit. Thus
although the term appears formally singular, the composite term is in
fact regular. Consequently, to evolve the fields directly
at~$\mathscr{I}^+$, it is necessary to calculate the limit of the
whole system of equations as~$r\rightarrow r_{\mathscr{I}}$. (Similar
calculations performed in full GR, rather than models, can be found
for instance in~\cite{MonRin08}). From the definition of~$R(r)$, one
finds that
\begin{align}
\frac{R'}{R^2} \rightarrow 2. \nonumber
\end{align}
The limit that has to be taken carefully is~$R'\Psi^-/R$ when~$p=1$,
which is the case for the~$U$ and~$F$ wave operators. To calculate
this limit, we use the l'Hospital's rule, giving
\begin{equation*}
\frac{R'}{R}\Psi^- \rightarrow -\p_r\Psi^- \,. 
\end{equation*}

\section{Results}
\label{numerics}

\begin{figure*}[t]
   \includegraphics[scale=0.36]{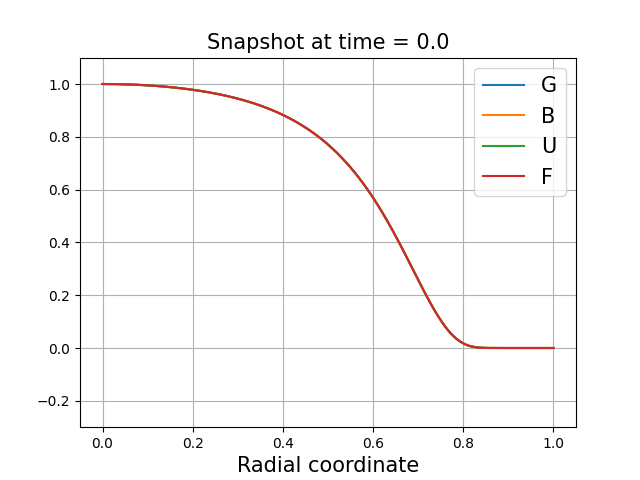} 
   \includegraphics[scale=0.36]{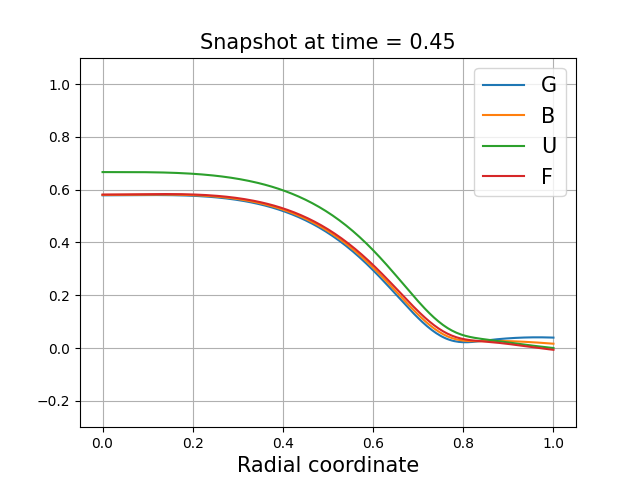} 
   \includegraphics[scale=0.36]{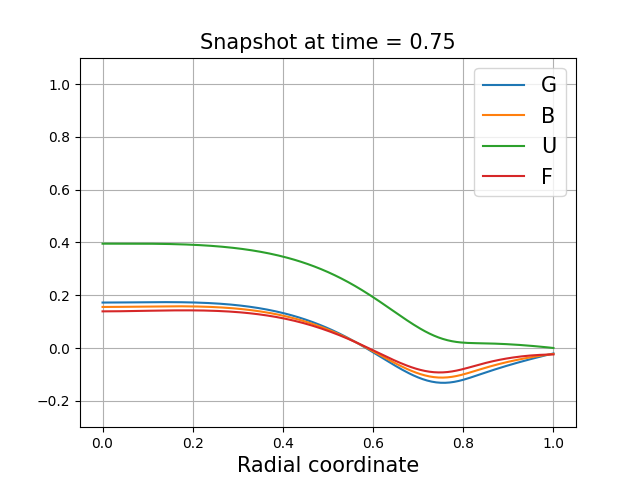} \\
   \includegraphics[scale=0.36]{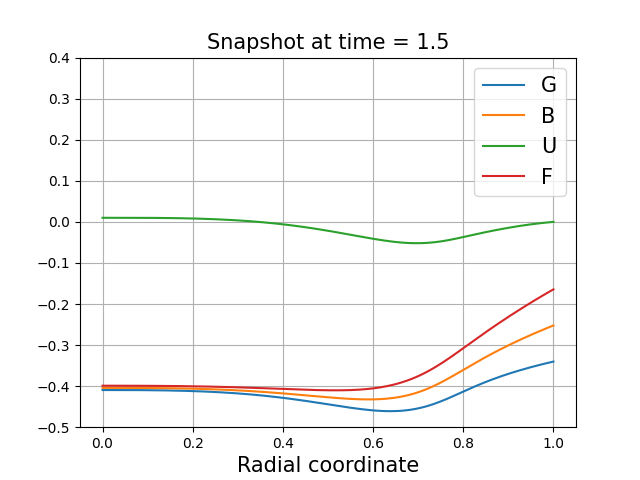} 
   \includegraphics[scale=0.36]{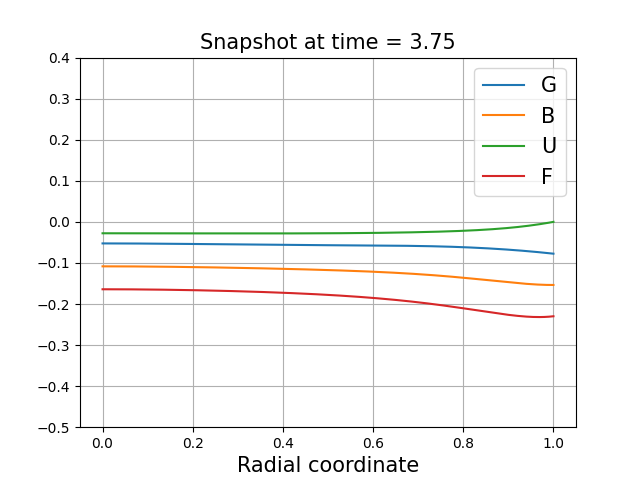} 
   \includegraphics[scale=0.36]{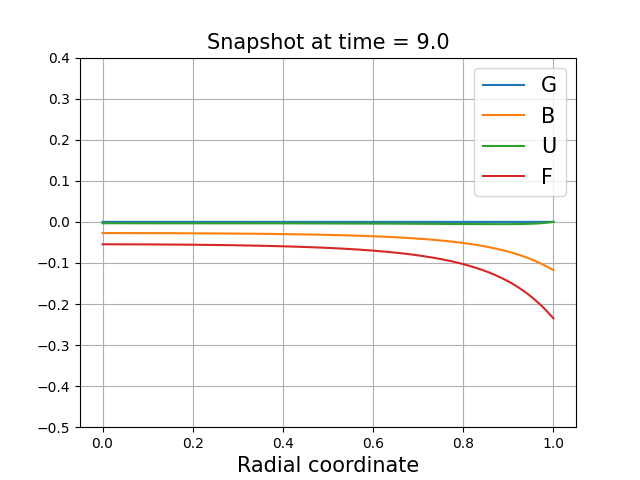} 
   \caption{Snapshots of spherically symmetric evolutions. Observe
     that all the fields reach a stationary state but have different
     asymptotics. In particular, the good field leaves entirely the
     domain and the ugly field is zero at~$\mathscr{I}^+$ for all
     times. \label{spherical_snapshots}}
\end{figure*}

We have implemented the GBUF model, employing the compactified
hyperboloidal coordinates~$(t,r,\theta^A)$ described above, both
within a stand-alone code that explicitly assumes spherical symmetry,
and within the 3d NRPy+ infrastructure. The code developed from this
infrastructure can be obtained at~\cite{CPBgit}.  Both implementations
use the method of lines with a fourth-order Runge-Kutta method for
time integration. In the spherically symmetric case, the grid has
points lying exactly at the origin and at~$\mathscr{I}^+$, as opposed
to the 3d case, in which a staggered grid is used. Spatial derivatives
are approximated by second-order accurate centered finite differences,
except at infinity, where one-sided derivatives are taken following a
truncation error matching approach (see~\cite{GauVanHil21}).

In both codes, the interior and outer boundaries require special
treatment. The interior boundary corresponds to the boundary of the
angular coordinates, in the~$3$d case or to the origin~$r=0$, in all
cases. Ghost points in the~$r<0$ case are populated using the parity
of the original fields. The original variables~$g$,~$b$,~$u$ and~$f$
are even in~$R$. In the 3d implementation, angular ghost points are
filled with the values of the corresponding gridpoints inside the
domain. In the continuum problem, no boundary condition is needed
at~$r=r_{\mathscr{I}}$, and we find it sufficient to fill
the~$r > r_{\mathscr{I}}$ ghost points with extrapolation, which we
take to be fourth order.

It is well known that the origin is a coordinate singularity in
spherical polar coordinates. This fact can be problematic in numerics
if the most naive discretizations are used, and turns out to be
particularly subtle for first order systems. Because of this, Evans
method~\cite{Eva84, GunGarGar10} is used to treat the equations at the
origin. To do this, we first use the
identity~$\frac{R'}{R} \Psi = \left( \frac{1}{r} -
  \frac{\Omega'}{\Omega} \right) \Psi$. We then
discretize~$\p_r \Psi + \frac{2}{r} \Psi$ by
\begin{align*}
  \left[ \p_r\Psi+\frac{2}{r}\Psi \right]_j
  \approx 3\, \frac{r_{j+1}^2\Psi_{j+1}-r_{j-1}^2\Psi_{j-1}}{r_{j+1}^3-r_{j-1}^3} \, .
\end{align*}
This trick is used to replace all the terms in which~$R^{-1}$ appears
explicitly as a coefficient of an evolved field.

Finally, we apply artificial dissipation on all variables on the whole
grid. For the spherically symmetric runs, this dissipation is just the
standard fourth order Kreiss-Oliger dissipation~\cite{KreOli73} with a
dissipation parameter~$\sigma=0.02$ using Evans method at the origin,
and instead~$\sigma = 0.01$ with our~SBP inspired discretization
(described in detail in section~\ref{Sec:Spherical}). In NRPy+,
because of the use of spherical polar coordinates, the dissipation
operators are adjusted with~$r^{-1}$ and~$(r\sin\theta)^{-1}$ factors
on the~$\theta$ and~$\phi$ derivatives,
respectively~\cite{MewZloCam18, RucEtiBau18}. Nevertheless, in the
full~3d case, more dissipation is required to eliminate high frequency
errors coming from the origin that eventually make the simulations
fail. This noise is likely due to the fact that the coordinate
singularity is more severe without symmetry, as it affects all points
with~$\theta = 0, \pi$. In this setting, following~\cite{MewZloCam18,
  RucEtiBau18}, we therefore use an overall dissipation factor of
around~$\sigma=0.4$. By performing evolutions on plain Cauchy slices,
we have verified that these modifications are not particular to the
use of hyperboloidal coordinates.

As is standard for error analysis in numerical work, we perform
norm-convergence tests for both setups. Because of the change of
coordinates and the introduction of our slightly non-standard
first-order reduction variables, the energy norm associated with the
plain scalar field takes the form~\cite{GauVanHil21}
\begin{align}\label{3Dnorm} 
  E(t) &= \int \left[ \left( \frac{2R'-1}{2R'\chi^2}
                (\Psi^+-\chi'\Psi)^2 \right)
                + \frac{1}{2R'}(\Psi^-)^2 \right. \nonumber \\
              &\qquad \left.
                 +\frac{1}{R^2}(\Psi_\theta^2+\Psi_\phi^2) \right]
                \frac{R'R^2}{\chi^2} \sin(\theta)drd\theta d\phi\,. 
\end{align}  
This is the norm used in our numerical convergence tests.

\subsection{Spherically symmetric test bed for 3d evolutions}
\label{Sec:Spherical}

Many interesting features of the model can already be studied in the
spherically symmetric case, for which the evolutions are performed
with the stand-alone infrastructure used for instance
in~\cite{VanHusHil14,GasGauHil19}. We give centered Gaussian~ID
corresponding to~$\psi(t=0,r) \propto e^{-R(r)^2}$
and~$\p_T \psi(t=0,r)=0$ for all the fields, and define the~ID for all
the~FOR variables accordingly.

One feature of interest is the rate of decay of the fields near
null-infinity. Placing certain assumptions on initial data and
employing the asymptotic systems approach of~\cite{LinRod03}, specific
rates were predicted in~\cite{DuaFenGas22}. The snapshots in
figure~\ref{spherical_snapshots} show that the evolved fields
remain~$O(1)$ for all times in accord with these predictions. Since we
choose the~ID for the ugly fields corresponding to~$m_u = 0$, the ugly
fields vanish at~$\mathscr{I}^+$. The rest of the fields behave
asymptotically like the~$G$ variables and so oscillate
at~$\mathscr{I}^+$. Interestingly near~$\mathscr{I}^+$, the~$B$
and~$F$ fields appear to reach a stationary but non-vanishing state at
late (hyperboloidal) times.

It is entirely expected that there is no incoming signal
from~$\mathscr{I}^+$, but it is interesting that this behavior is
captured well by the numerical approximation, and furthermore that we
see no evidence of incoming waves being generated as reflections in a
neighborhood of~$\mathscr{I}^+$ either. Instead we see the signal
practically leaving the domain in finite hyperboloidal time.

\begin{figure*}[t] 
  \includegraphics[scale=0.5]{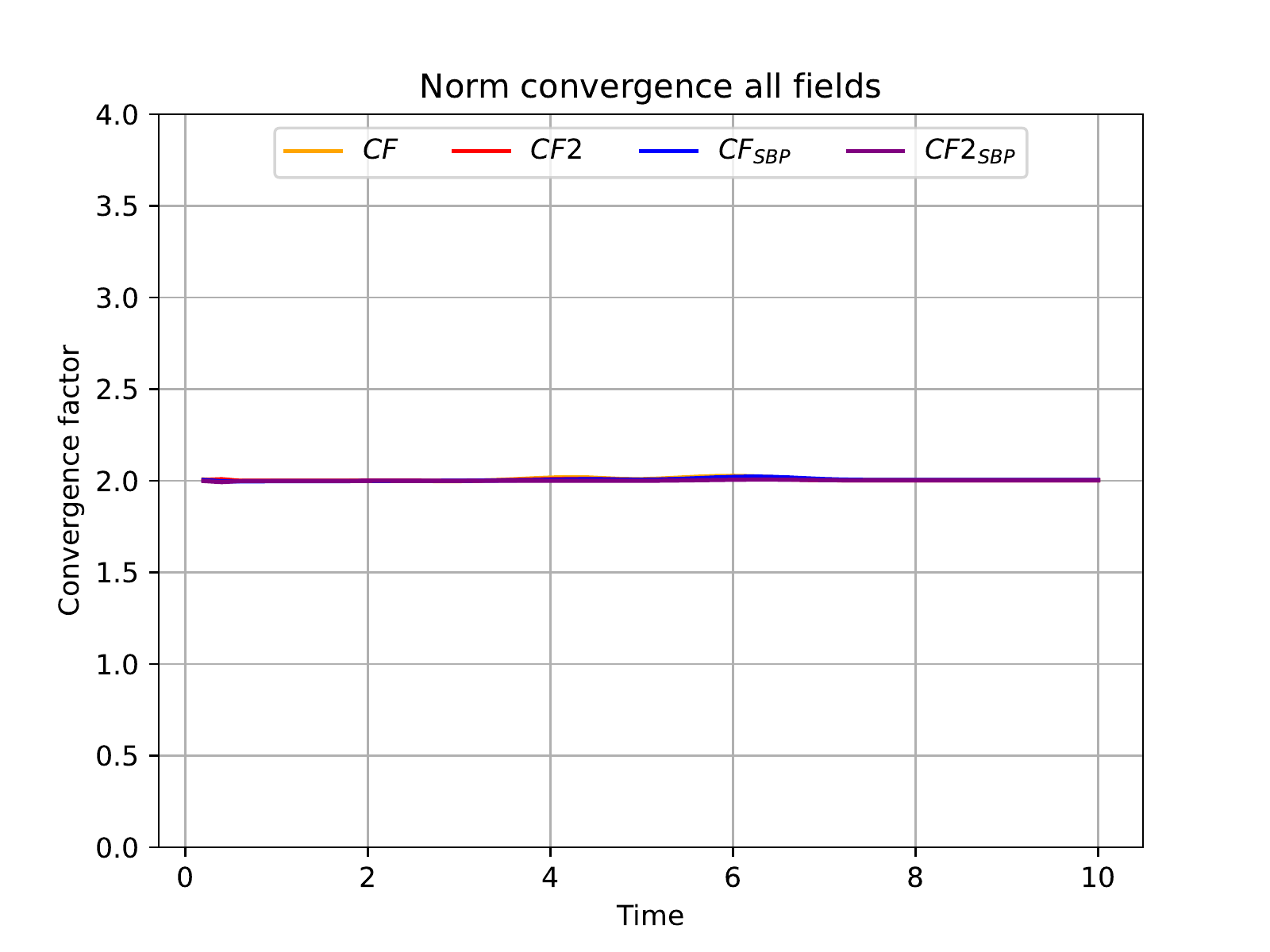}
  \includegraphics[scale=0.5]{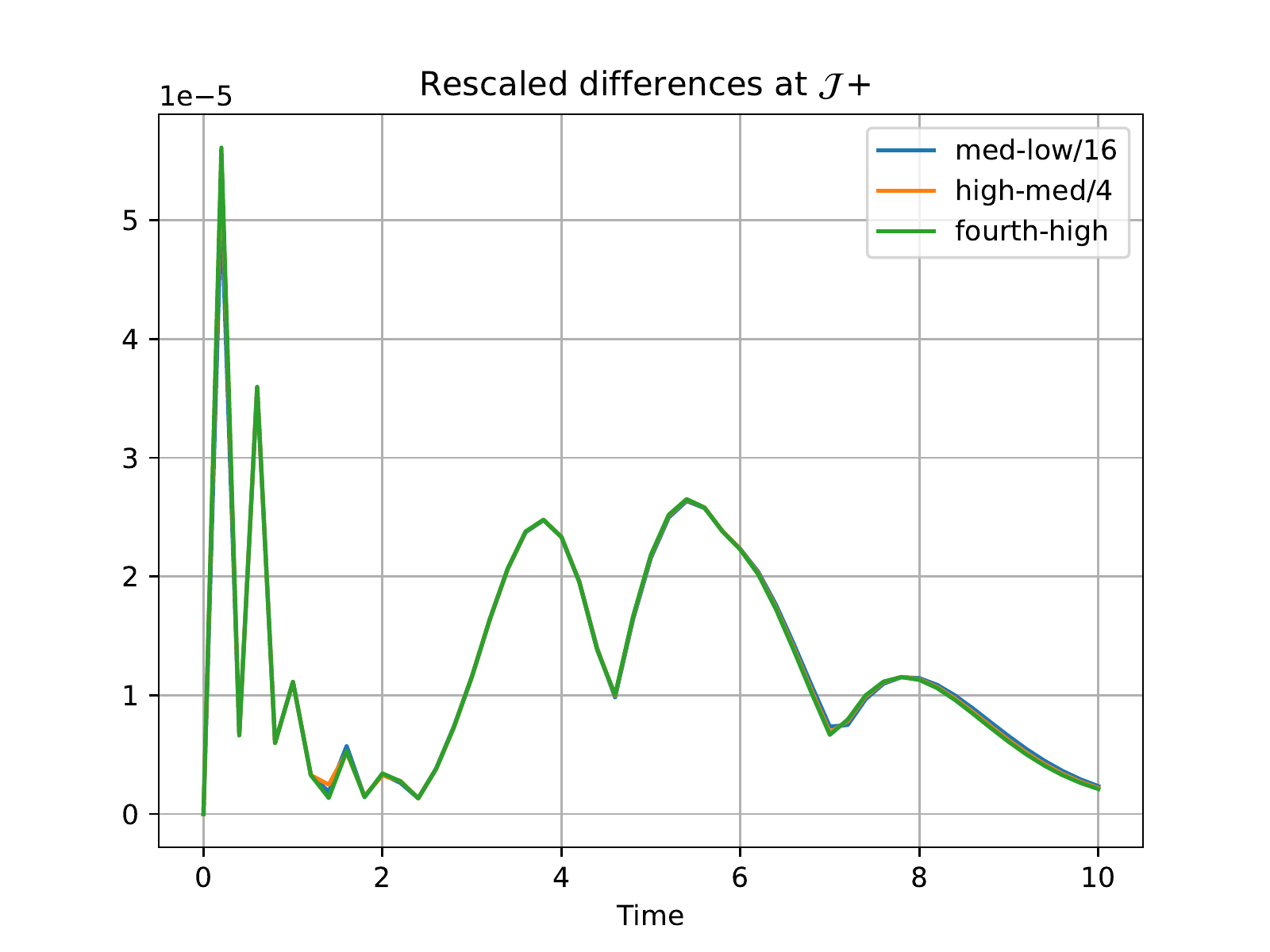}
  \caption{In the left panel we show norm convergence for the entire
    state vector. Orange and red curves show the convergence factor
    for standard finite differences with the Evans method for first,
    second, third and second, third and fourth resolutions
    respectively. Blue and purple curves show the analogous norm
    convergence factor for the simulations with our second scheme. The
    data display robust second order convergence. The right panel
    shows the sum over all the fields, including the reduction
    variables, of the absolute value of the rescaled differences
    at~$\mathscr{I}^+$ from the Evans discretization. The three
    curves overlap very well at all times, meaning we get excellent
    pointwise convergence
    at~$\mathscr{I}^+$.\label{fields_convergence}}
\end{figure*}

Next, we successfully performed long convergence tests on our
numerical solutions, examining the data both pointwise and in the
energy norm~\eqref{3Dnorm}. At the base resolution, we took~$200$
gridpoints in the radial coordinate, doubling resolution at every
level of the convergence test. The energy norm convergence rate,
computed on all the fields, is shown as a function of time in
figure~\ref{fields_convergence}.

Recall that the main objective of this work is to show that the system
at hand, viewed as a model for the~EFEs in~GHG, can be reliably
numerically evolved on compactified hyperboloidal slices. We therefore
check pointwise convergence for the gridpoints {\it
  at}~$\mathscr{I}^+$ which, in the spherical case, corresponds to a
single gridpoint at each time. Rescaled differences at this point as a
function of time are shown in figure~\ref{fields_convergence}. Since
the three curves lie on top of each other, we get the expected
convergence order.

In all of the above spherical runs, we employed the standard Evans
method as described above. To demonstrate that our numerical results
are not strongly dependent on this particular choice of variables and
numerical scheme, we performed spherically symmetric evolutions also
using another discretization. This scheme is similar to the~SBP
discretization derived in~\cite{GauVanHil21}. We use a similar (though
not identical) discretization as we still do not have an~SBP scheme
for the linear part of whole~GBUF model. The derivation of such a
scheme is left for future work. Both the Evans and~SBP-like
discretization have the property that they stabilize the approximation
even in the absence of artificial dissipation, although dissipation
does help suppress undesirable error at the origin.

To implement the~SBP-like discretization, we define the rescaled
fields as
\begin{align}\label{good_rescaling}
  \Psi \equiv \chi \psi \, , \, \Psi^+ \equiv \chi^2 \psi^+ \, , \,
  \Psi^- \equiv \chi \psi^- \, ,
\end{align}
where, as before,~$\psi$ is one of the~$g$,~$b$,~$u$ or~$f$ variables
and~$\psi^+$ and~$\psi^-$ are defined in~\eqref{FOR}. The
associated~FOR constraints are
\begin{align}
  \frac{\chi}{R'} \left( \p_r \Psi + \frac{\Psi^-}{2} \right)
  - \Psi \chi' + \left( \frac{1}{2 R'} - 1 \right) \Psi^+ = 0 \, .
\end{align}
Defining
\begin{align}
\tilde{\p}_r \Psi \equiv (\chi^2 R^{-2}) \p_r (R^2 \chi^{-2} \Psi) \, ,
\end{align}
the equations take the form
\begin{align}\label{SBP_G}
\p_t \Psi &=  \frac{1}{2} \left(\frac{\Psi^+}{\chi} + \Psi^- \right) \, , \nonumber \\
  \p_t \Psi^+ &=  \frac{\chi}{2(2R'-1)} \left( (\p_r + \tilde{\p}_r)
                  \left(\chi^{-1} \Psi^+ \right) + (\p_r - \tilde{\p}_r) \Psi^- \right) \nonumber \\
& - \frac{R'}{2R'-1} \left( \frac{p \Psi^+}{\chi} + (\chi' + p) \, \Psi^- + S_\Psi \right) \, , \nonumber \\
\p_t \Psi^- = & \frac{1}{2} \Big( (\tilde{\p}_r - \p_r) \left(\chi^{-1} \Psi^+ \right) - (\tilde{\p}_r + \p_r) \Psi^- \Big) \nonumber \\
& + \frac{R'}{\chi} \left( \frac{(\chi'-p) \, \Psi^+}{\chi} - p \Psi^- - S_\Psi \right) \, , \,
\end{align}
in the bulk,
\begin{align}\label{SBP_G_origin}
  \p_t \Psi = \Psi^+ \, , \,
  \p_t \Psi^+ = \p_t \Psi^- = 3 \p_r \Psi^+ - 2 p \Psi^+ - S_{\Psi_0} \, ,
\end{align}
at the origin. Here, the source terms~$S_G$ and~$S_U$ are identically
zero, and
\begin{align}
S_B = & \frac{1}{2} \left( F^- + \frac{F^+}{\chi} \right) + \frac{1}{4} \left(\frac{G^+}{\chi} + G^- \right)^2 \, , \nonumber \\
S_F = & \frac{1}{2} \left(\frac{G^+}{\chi} + G^- \right)^2 \, ,
\end{align}
with their limits at the origin being
\begin{align}
S_{B_0} = F^+ + (G^+)^2 \, , \, S_{F_0} = 2 (G^+)^2 \, .
\end{align}
The limits at~$\mathscr{I}^+$ are straightforward. The equations
at~$\mathscr{I}^+$ are obtained by taking the limits
\begin{align*}
  & U^- \rightarrow 0 \, , \, F^- + \frac{(G^-)^2}{2}
    \rightarrow 0 \, , \, \frac{R' U^-}{R} \rightarrow - \p_r U^- \, , \nonumber \\
  & \frac{R'}{2R} \left( F^- + \frac{(G^-)^2}{2} \right) \rightarrow
    - \p_r F^- - G^- \p_r G^- \, .
\end{align*}
In our implementation we compute them numerically without using
the~$\tilde{\p}_r$ operator.

We perform our convergence tests for this discretization in the norm
\begin{align}\label{SBP_Norm}
  E(t) = \int_0^{r_\mathscr{I}} \left( \frac{1}{2} r^2 \Psi^2
  + \varepsilon_S \right) \, dr \, ,
\end{align}
with
\begin{align}\label{good_energy_norm}
  \varepsilon_S = \frac{1}{2} \left[ \frac{(\Psi^-)^2}{2}
  + \left( \frac{2R'-1}{2 \chi^2} \right) (\Psi^+)^2 \right]
  \frac{R^2}{\chi^2} \, ,
\end{align}
where~$\psi$ stands for any of the~$g$,~$b$,~$u$ or~$f$ variables,
and~$\Psi$,~$\Psi^+$ and~$\Psi^-$ are defined as
in~\eqref{good_rescaling}. To make the norms in the two discretization
schemes compatible with each other, we add~$r^2 \Psi^2/2$ to the
integrand of the norm~\eqref{3Dnorm}, when working in spherical
symmetry. However, no such modification is introduced in the full~$3$d
case, and the norm~\eqref{3Dnorm} is used. In all cases, the argument
in~$E(t)$ stands for the hyperboloidal time over which the integral on
the right is evaluated.

As can be seen in Figure~\ref{fields_convergence}, all the results in
the~SBP inspired discretization look similar to those obtained using
Evans method. The~FOR constraint violations in both the schemes also
converge perfectly at second order. These violations arise only
because of the discretization, as our~ID satisfies the reduction
constraint in the continuum limit.

\begin{figure*}[t!]
   \centering
   \includegraphics[scale=0.305]{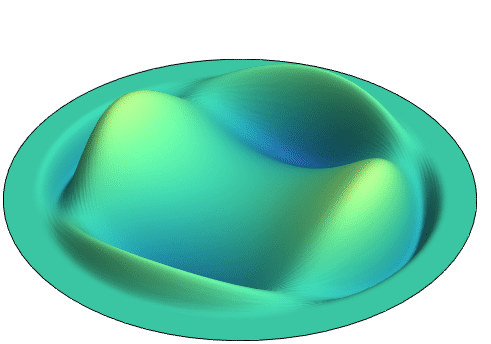} \quad
\hspace{-0.1cm}\includegraphics[scale=0.305]{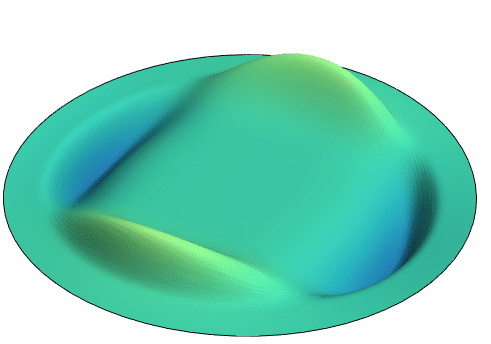} \quad
\hspace{-0.1cm}\includegraphics[scale=0.305]{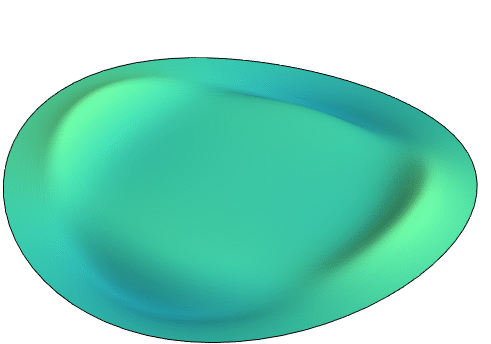} \quad
\hspace{-0.1cm}\includegraphics[scale=0.305]{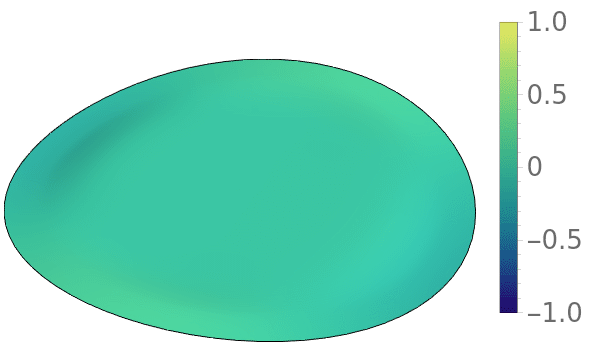} \quad
  
   \includegraphics[scale=0.305]{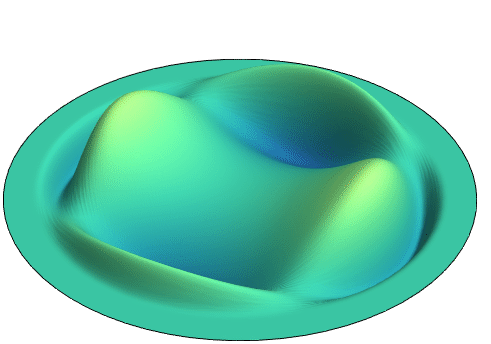} \quad
\hspace{-0.1cm}\includegraphics[scale=0.305]{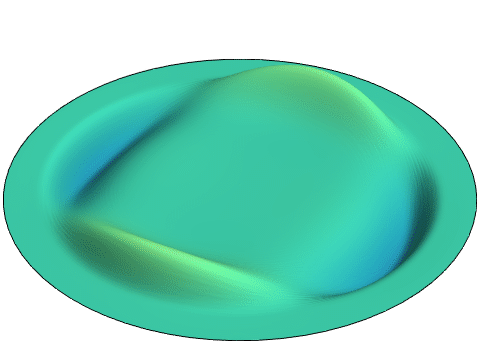} \quad
\hspace{-0.1cm}\includegraphics[scale=0.305]{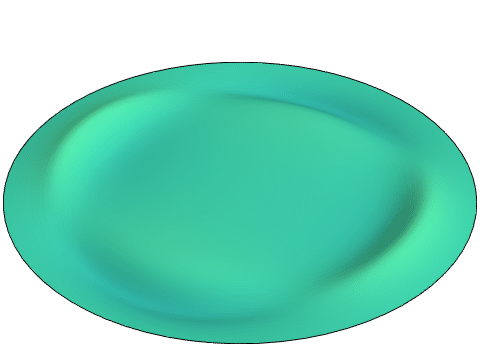} \quad
\hspace{-0.1cm}\includegraphics[scale=0.305]{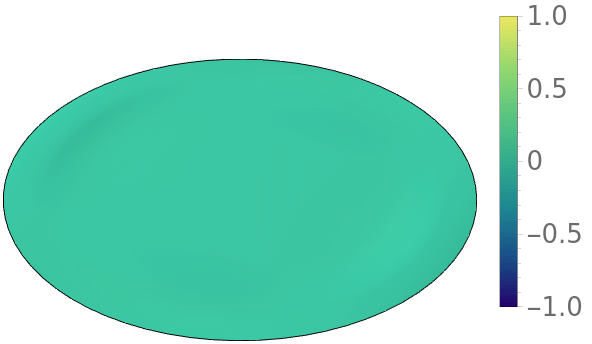} \quad
\caption{Snapshots of 3d evolutions of B and U fields (on the top and
  bottom rows) as functions of~$r$ and~$\phi$ for~$\theta\simeq1$
  radian. Observe that similar features can be seen as compared to the
  spherically symmetric runs, namely, a stationary solutions is
  reached and the ugly field is zero at~$\mathscr{I}^+$ for all
  times. The~$G$ and~$F$ fields look qualitatively similar.}
\label{3D_snapshots}
\end{figure*}

\subsection{Full 3d case}

Our next task is to see whether numerical evolutions of the~GBUF
system can be performed with null-infinity without symmetry
assumptions. To this purpose, we perform full 3d evolutions using
spherical polar coordinates. Unfortunately, even putting aside the
additional computational cost, this is not a completely
straightforward generalization of the previous case. The main
complication is that we now encounter a ``more singular'' behavior at
the coordinate singularity along the whole~$z$ axis.  To manage that
challenge, besides using Evans method, we follow the basic philosophy
of NRPy+, dividing the~$\phi$ derivative of the fields by $\sin\theta$
(see~\ref{FOR}). The corresponding dissipation operator also needs to
be modified as described above. Another strategy to manage these
challenges would be to make a multipatch approach, as, for instance,
in earlier numerics for the hyperboloidal treatment of the wave
equation with the pseudospectral \texttt{bamps}
code~\cite{HilHarBug16}. Presently, to evolve the full~$3$d system
numerically, we used~NRPy+~\cite{RucEtiBau18}, a numerical relativity
Python infrastructure that outputs optimized~C code for the
runs. Ideally, this~C code is just compiled and run
within~NRPy+. There were however two places where, in our
implementation, the C code was modified directly by hand because
the~NRPy+Python environment was not designed to automate the
particular bespoke changes we required. First were the~$r<0$
ghostpoints, which were populated by parity conditions. This is
because the~$\Psi^{\pm}$ fields evaluated at points~$r<0$
involve~$\Psi^{\mp}$ and~$\Psi$ evaluated at~$r>0$. Second was the
implementation of Evans method, which was not used (or needed) in the
original second order in space NRPy+ implementation of the wave
equation on Cauchy slices. All the modified files, including those
needed to create the code, together with instructions for compilation
and use of the code can be found at the aforementioned
link~\cite{CPBgit}.

To avoid directly facing the singular nature of the coordinates, the
code uses a staggered grid in the three spatial
coordinates,~$(r,\theta,\phi)$. This means that the spatial domain in
a coordinate, say~$r \in (0,1)$, is divided in~$N_r$ cells, and
there's a gridpoint in the center of each cell. The same is done for
the~$\theta$ and~$\phi$ domains. This is standard in the~NRPy+
infrastructure.

As a first test, we input the same~ID as in the spherical-code
with~$(200,4,4)$ gridpoints in the respective~$(r,\theta,\phi)$
coordinates at the base resolution. For this case we employed the same
dissipation operators and parameters as for the spherically symmetric
code, but with full 3d evolution, and made a side-by-side
comparison. For the same resolution, the difference in the
gridfunctions outputted from the different codes differ at most by
numbers of order~$O(10^{-3})$, for gridfunctions of order~$O(1)$, so
they differ by less than~$1\%$. Moreover, the largest differences
appear close to the origin, and for most of the grid the errors
decrease to as little as~$O(10^{-7})$. We also performed
norm-convergence tests for the spherical code with a staggered grid
and the 3d one, by tripling resolution 2 times in the radial
coordinate, and compared them directly. Norms calculated from the
different codes outputs differ by numbers of order~$O(10^{-3})$ for
all times, so they differ by less than~$1\%$ as well.  We take the
excellent agreement of the codes in this setup as evidence for the
correctness of the 3d implementation.

\begin{figure*}[t!] 
 \includegraphics[scale=0.5]{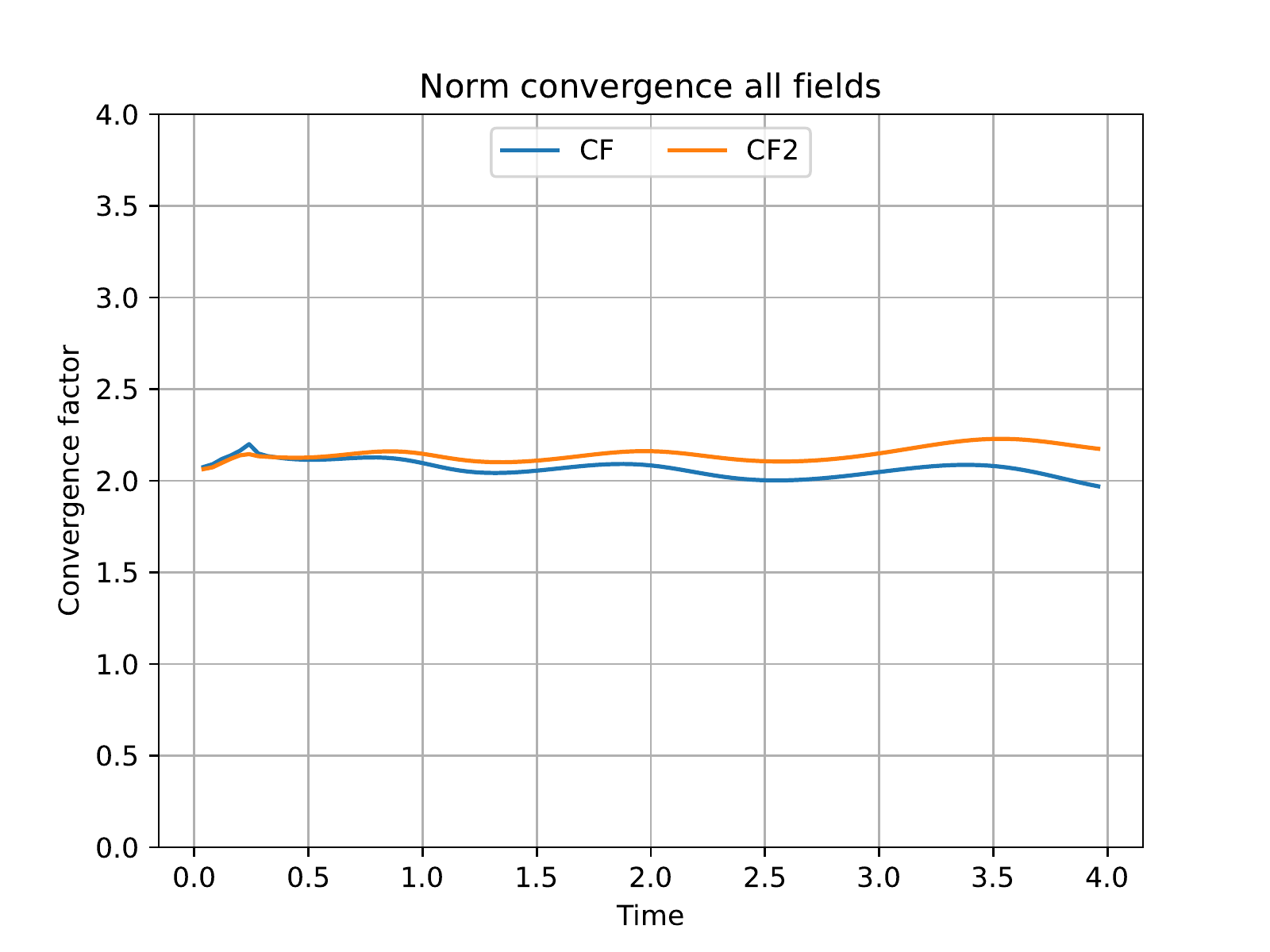}
 \includegraphics[scale=0.5]{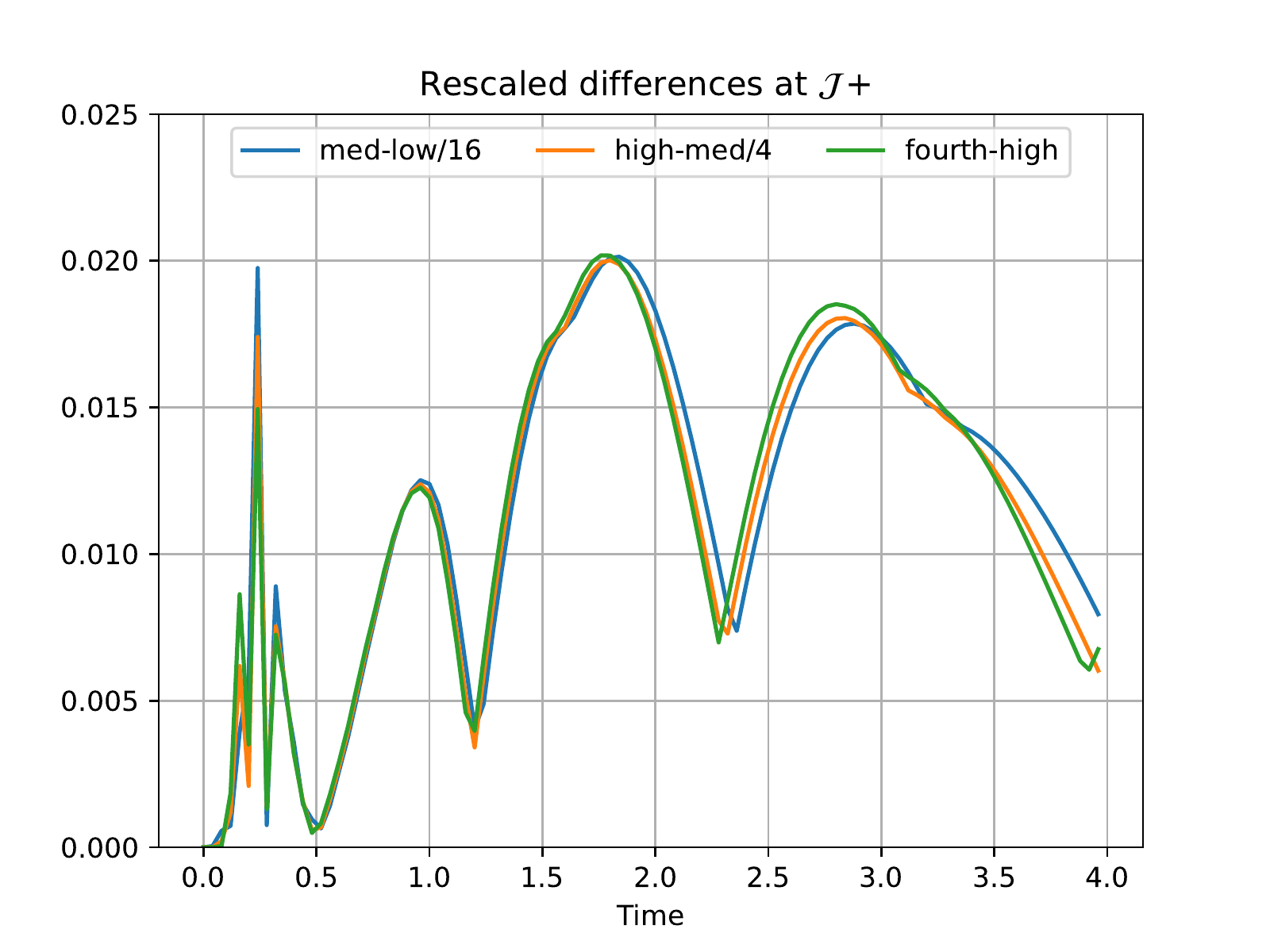}
 \caption{Norm convergence and absolute value of rescaled differences
   at~$\mathscr{I}^+$ for the state vector. We observe
   convincing norm convergence as a function of time with a slight
   drift for late times. Good pointwise convergence at~$\mathscr{I}^+$
   can also be seen from the figure on the right, even though these
   curves were calculated by extrapolating the data half a grid
   point.}
 \label{norm_convergence_3D}
\end{figure*}

For the ID with no symmetry, we use a partial-wave-like expression of
the form
\begin{align}
\psi(t=0) &= \frac{A}{4R} e^{-(1 + 4R)^2/16}\big[-4 -2R -15R^2 \nonumber \\
          &\quad  +8R^3 +16R^4 +e^{R}(4 -2R +15R^2\nonumber\\
          &\quad  +8R^3 -16R^4)\big]\,Y_{22}(\theta,\phi) \,,
\end{align}
where~$Y_{22}$ is the~$l=2,\, m=2$ spherical harmonic and the
amplitude parameter~$A$ is a constant. This choice was taken because a
smooth solution at the origin with no symmetry is needed for this
simulation, with the expression inspired by the d'Alembert
partial-wave solution to the wave equation for a given smooth
function~\cite{GunPriPul93, SuaVicHil20}, which in this case we take
to be~$S(R)=R^2 e^{ -(R-\frac{1}{4})^2 }$. We take the constant~$A=1$
for the bad, ugly and~$f$ fields and~$A=1/10$ for the~$g$ field. We
choose this because larger values of the~$g$ field make it completely
dominate the behavior in the~$b$ field through the nonlinearity, and
we wanted to see evidence of the linear radiation field besides.  In
figure~\ref{3D_snapshots} we show snapshots of the evolution of the
system as a function of~$r$ and~$\phi$ for a particular choice
of~$\theta$. The same features as in the spherically symmetric case
can be seen, namely, the appropriate decay rates and the signal
leaving the domain in finite hyperboloidal time, leaving behind a
stationary solution.

Norm convergence tests were also performed for the 3d setting, again
using the norm in expression~\eqref{3Dnorm}. We started
with~$(80, 16, 32)$ gridpoints in $(r,\theta,\phi)$ coordinates,
respectively, and increased resolution by a factor of~$1.5$ three
times in each coordinate. Fourth-order interpolation was used to match
the higher resolution gridpoints to the lowest one's grid, except at
the numerical boundaries, where the third order was used. The result
is shown in the left panel of
figure~\ref{norm_convergence_3D}. Respectable second order convergence
is visible in the plot. A slight drift from the ideal convergence rate
is seen for late times, although by these times, the data has already
mostly left the numerical domain (recall that a purely radially
outgoing pulse travelling at the speed of light takes around~$t=1$ to
reach~$r=r_{\mathscr{I}}$ in our coordinates).  Second order
pointwise convergence at~$r=r_{\mathscr{I}}$ is examined in the right
hand panel of figure~\ref{norm_convergence_3D}, and is similarly
promising.

\section{Conclusions}
\label{conclusions}

In this work, a direct extension of~\cite{HilHarBug16, GasHil18,
  GasGauHil19, GauVanHil21}, we made another step towards including
future null-infinity in generic numerical relativity simulations of
asymptotically flat spacetimes. Our strategy is to extend formulations
of GR, specifically the GHG system, that are known to work well in the
strong-field region on to compactified hyperboloidal slices. One
challenge is that, unlike the Conformal Field Equations~\cite{Fri81a},
the resulting system of PDEs may not be regular at null-infinity
because, depending on the specific choice of gauge source function,
non-principal terms may lack sufficient decay in radius to offset
divergences due to the radial compactification. Our direct starting
point here was the demonstration provided by~\cite{DuaFenGas22} that,
with care, the gauge source functions can be chosen so that the
worst~$O(\log(R)/R)$ decay present with a naive choice of the gauge
sources is circumvented. Under that approach, the evolved variables
have full~$O(1/R)$ decay near~$\mathscr{I}^+$. The subtlety that
remains to be handled numerically are {\it formally singular}
terms. The purpose of this paper was to demonstrate that these terms
could be managed numerically both in spherical symmetry and full 3d
simulations. Since this difficulty can be made explicit without the
full complication of the EFEs, we studied a toy model. We presented
numerical evolutions of the~\textit{Good-Bad-Ugly-F} model, a system
of nonlinear wave equations that mimics the asymptotic properties of
the~EFEs in~GHG. We showed that the various different fields, each
with different decay rates towards~$\mathscr{I}^+$, can be evolved
numerically on compactified hyperboloids.

In related earlier work~\cite{HilHarBug16} pseudospectral numerics for
the wave equation on hyperboloidal slices was presented. Here instead,
because of the potentially slow asymptotic decay of fields in the GBUF
model, purely as a proof-of-principle, we employed finite differences
exclusively for the approximation of spatial derivatives. We made two
numerical implementations, one in explicit spherical symmetry (with
two distinct choices of variables and discretization), the other in 3d
using NRPy+. Our first important result was that the decay rates for
each of the fields predicted in~\cite{DuaFenGas22} were reliably
obtained in both codes, and for all of the initial data sets we
treated. Moreover, the expected properties of hyperboloidal evolution
of wave-like equations with our setting were observed. Among these
were the finite but non-vanishing velocities of the signals going
through~$\mathscr{I}^+$ and, for a subset of the fields, the presence
of non-vanishing near-stationary solutions after a finite
hyperboloidal time. Clean norm convergence and pointwise convergence
at~$\mathscr{I}^+$ is seen for all cases from our numerics. In
summary: the results presented here show that there should be no
problem in managing the asymptotic properties of the~EFEs in~GHG on
compactified hyperboloidal slices.

An interesting point that we have made no attempt whatsoever to
understand here is the effect of choosing initial data that decay only
very slowly towards~$\mathscr{I}^+$. This would make direct contact
with~\cite{DuaFenGas23}, where such initial data were considered for a
subsector of the GBUF model. In the context of GR, such data has
relevance to the question of the peeling property at~$\mathscr{I}^+$
(see~\cite{Lin17,GasVal18,Fri18,Keh21} for recent work in this
direction). Unfortunately, treating such initial data would force a
complete rethink of the numerical strategy.

Many pieces of the puzzle are now in place for a~$3+1$ implementation
of the EFEs on hyperboloidal slices. There are a number of outstanding
questions however, including for instance our incomplete understanding
of charges at~$\mathscr{I}^+$ expressed in generic generalized
harmonic gauges, the lack of a clear (stand-alone) local
well-posedness theory on hyperboloidal slices and the hands-on
construction of initial data for a variety of scenarios of
interest. Progress on all these fronts is expected. In the near-term
we will present a comprehensive set of spherical numerics for full GR.

\acknowledgments

The Authors wish to thank Miguel Duarte, Edgar Gasperin and Thanasis
Giannakopoulos for helpful discussions and especially for their
continuing collaboration. We also greatly profited from interaction
with Leonardo Werneck and Zachariah Etienne and their introduction to
NRPy+. SG thanks Sascha Husa for local hospitality and travel support
at~UIB, where part of this work was performed. Part of the
computational work was performed on the Sonic cluster at ICTS. The
authors thank FCT for financial support through Project
No. UIDB/00099/2020 and IST-ID through Project
No. 1801P.00970.1.01.01.  S.G.’s research was supported by the
Department of Atomic Energy, Government of India; Infosys-TIFR Leading
Edge Travel Grant, Ref. No.: TFR/Efund/44/Leading Edge TG (R-2)/8/;
and by the Ashok and Gita Vaish Early Career Faculty Fellowship owned
by Prayush Kumar at the International Centre for Theoretical Sciences.

\bibliography{GBUF.bbl}

\end{document}